\documentclass[12pt]{iopart}

\usepackage{iopams}  
\usepackage{graphicx}
\usepackage{color}
\usepackage{lscape}
\begin{document}

\newif\ifRR
\RRtrue

\title[Using the Quantum Zeno Effect for Suppression of Decoherence]
{Using the Quantum Zeno Effect for Suppression of Decoherence} 

\author{Yasushi Kondo}
\address{Department of Physics, Kinki University, Higashi-Osaka 577-8502, Japan}
\ead{ykondo@kindai.ac.jp}

\author{Yuichiro Matsuzaki}
\address{NTT Basic Research Laboratories, NTT Corporation, Kanagawa 243-0198, Japan}
\author{Kei Matsushima}
\address{Department of Physics, Kinki University, Higashi-Osaka 577-8502, Japan}
\author{Jefferson G. Filgueiras} 
\address{Fakult\"at Physik, Technische Universit\"at Dortmund, Otto-Hahn-Strasse 4, 
D-44221 Dortmund, Germany} 
\address{Instituto de F\'isica de S\~ao Carlos, 
Universidade de S\~ao Paulo, P.O. Box 369, S\~ao Carlos, 13560-970 SP,
Brazil}

\vspace{10pt}
\begin{indented}
\item[]August 2015
\end{indented}

\begin{abstract}
Projective measurements are an essential element of quantum mechanics.
In most cases, they cause an irreversible change of the quantum system 
on which they act. However, measurements can also be used to stabilize 
quantum states from decay processes, which is known as the Quantum Zeno 
Effect (QZE). Here, we demonstrate this effect for the case of a superposition 
state of a nuclear spin qubit, using an ancilla to perform the 
measurement. As a result, the quantum state of the qubit is protected 
against dephasing  without relying on an ensemble nature of NMR 
experiments. We also propose a scheme to protect an arbitrary 
state by using QZE. 
\end{abstract}

%
\vspace{2pc}

%
\submitto{\NJP}
%
%
%

\section{Introduction}

Interactions between a quantum system and its environment lead to
changes of the system state and 
can result in the loss of coherence. This effect, often called
decoherence, is a limiting 
factor for many applications, such as quantum computing. 
The loss of information associated with this process can be measured 
by the decreasing overlap between
the initial and current states. This overlap (state fidelity) changes 
quadratically for times short enough
compared to the inverse of the strength of the system-environment coupling  
and the correlation time of the environment, provided this coupling
is also static on the relevant timescale \cite{nakazato,palma}.
If such a system is repeatedly projected back to the original state by
measurements performed before the state has changed significantly, the
effect  
of the system-environment interaction can be effectively 
eliminated in the limit of sufficiently frequent measurements.
This is termed the Quantum Zeno Effect (QZE) \cite{misra}. There are many 
potential applications of QZE, 
like entanglement generation 
\cite{nakazatoyuasa}, quantum metrology \cite{matsuzakisimonjoe,chin}, and 
quantum imaging \cite{kwiat}.

It is worth mentioning that QZE is different from dynamical decoupling, 
which also suppresses decoherence. Although both schemes were unified
from the viewpoint of a strong interaction between
a system of interest and the others
\cite{fachi}, there is a clear difference between them.
While dynamical decoupling uses unitary control to average out the 
system-environment interaction \cite{lidar}, the non-unitary dynamics 
introduced by the measurements is essential for the state stabilization 
in QZE.

In the simplest case, the evolution suppressed by QZE is driven by a
single interaction to a static external degree of freedom.
Typical examples include a transition between two states of a single trapped
ion \cite{itano}, nuclear magnetic resonance (NMR) \cite{nmr}, and atomic
systems \cite{atomtwo,atomthree}. Another example is the confinement of a unitary
dynamics in a specific subspace, as demonstrated in a rubidium
Bose Einstein condensate \cite{atomsubspace}. In a more sophisticated case, QZE 
restrains the non-unitary dynamics of an open quantum system \cite{misra}, 
instead of suppressing a unitary evolution driven by external fields.
One experiment controlled such non-unitary dynamics, the escape
rate of atoms from a trap, using the QZE \cite{atomone}. 
However, QZE to suppress the dephasing of a two-level
system was not demonstrated yet.

In this paper, we present a proof-of-principle demonstration that  
QZE is equally possible to stabilize the state of a 
quantum system subject to dephasing by successive measurements. 
We employ a liquid-state NMR Quantum Computer
with a two-spin molecule. However, the ensemble nature of NMR is not essential 
to implement the measurements in our experiments, unlike 
in \cite{nmr,antizenonmr}. 
A similar scheme was proposed for a non-ensemble system 
in \cite{zenoyuichiro}. Also, QZE experiments 
without relying on an ensemble average were reported \cite{atom_single}. 

This paper is organized as follows. 
In Sec.\ II, a brief review on QZE and a detailed theoretical background of 
our experiment are presented. 
We show a proof-of-principle demonstration, including a characterization
of the molecule employed in our experiments, in Sec.\ III. 
The last section is devoted to conclusion and discussions, where we 
propose a scheme to protect an arbitrary state by using QZE.

\section{Theory}
\subsection{Quantum Zeno Effect}
In order to understand the essence of QZE, 
it is sufficient to consider 
a coherent superposition of a qubit
being dephased by a randomly fluctuating classical 
field \cite{nakazato,palma}, with
the concept of ``mixing process'' \cite{kondo_a_rel}. 
The relevant Hamiltonian is thus
\begin{eqnarray}\label{1}
\mathcal{H}(t)&=&H_s+H_I+H_{en}, \\
 H_s&=&\frac{\omega _{s}}{2}\sigma _z, \\
 H_I&=& \lambda \sigma_z \hat{A}, \\
 H_{en}&=&\omega _{en}\hat{B}, 
\end{eqnarray}
where $\omega _s$ denotes the energy of the system, $\lambda$ 
denotes the strength of the fluctuating
field, $\hat{A}$ denotes the environmental operator to represent the
interaction with the system, $\omega_{en}$ denotes the environmental energy, 
$\hat{B}$ denotes the environmental operator,
and $\sigma_i$ is a Pauli matrix.
The Hamiltonian 
contains only $\sigma_z$ because only pure dephasing is considered.
We go to an interaction picture, obtaining
\begin{eqnarray}
 H_I(t)\simeq \lambda \sigma_z \hat{A}(t),
\end{eqnarray}
where $\hat{A}(t) =e^{i\omega _{en}\hat{B}}\hat{A}e^{-i\omega _{en}\hat{B}}$.

We prepare a superposition state $\displaystyle \vert + \rangle =\frac{1}{\sqrt{2}}(\vert
0\rangle + \vert 1\rangle)$ for the system. The density matrix of the system and
the environment is 
$\displaystyle \rho(0) = \rho_s(0) \otimes \rho _{en}$ 
where  $\rho_s(0) = \vert + \rangle \langle + \vert$ and 
$\rho_{en}$ denotes the initial state of the environment.
Formally integrating the von Neumann equation $\dot\rho=-i[\mathcal{H}, \rho]$,
we obtain
\begin{eqnarray*} 
\rho(t) = \rho(0) - i\lambda\int_{0}^{t}dt_{1}[\sigma_z\hat{A}(t_1) ,\rho(t_1)].  
\end{eqnarray*}
The natural unit system with $\hbar = 1$ is used. 
Iterating $\rho(t_1)$ on the right hand side and
replacing
$\rho(t_2)$ by $\rho(0)$, 
we have the second order expression
\begin{eqnarray}
 \label{4}
\rho(t) &\simeq& \rho(0) -
i\lambda\int_{0}^{t}dt_{1}[\sigma_z\hat{A}(t_1) ,\rho(0)]
\nonumber \\  &-&
\lambda^{2}\int_{0}^{t}\int_{0}^{t_1}dt_{1}dt_{2}
[\sigma_z \hat{A}(t_1),[\sigma_z\hat{A}(t_2), \rho(0)]].\nonumber
\end{eqnarray}
Tracing out the environment, we get
\begin{eqnarray}
\label{eq:td_rhos}
\rho _s(t) 
&=&\rho _s(0)-\lambda^{2}\int_{0}^{t}\int_{0}^{t_1}dt_{1}dt_{2}
\langle \hat{A}(t_1)\hat{A}(t_2)\rangle [\sigma_z,[\sigma_z, \rho_s(0)]].
\end{eqnarray}
where $\rho _s(t) ={\rm Tr}_{en}[\rho (t)]$ denotes the density operator of
the system, ${\rm Tr}_{en}[\cdots]$ denotes a partial trace of the environment, 
and $\langle \hat{A}(t_1)\hat{A}(t_2)\rangle ={\rm Tr}[\hat{A}(t_1)\hat{A}(t_2)
\rho_{en}]$ denotes a correlation function.
Here, we assume 
an unbiased noise $\langle \hat{A}(t)\rangle ={\rm Tr}[\hat{A}(t)\rho
_{en}]\simeq 0$ and a time symmetry $\langle \hat{A}(t_1)\hat{A}(t_2)\rangle
=\langle \hat{A}(t_2)\hat{A}(t_1)\rangle $.

If the correlation time of the noise is much shorter than the relevant time scale 
of the system \cite{c_time_delta}, the duration of each measurement 
in  our case, 
we can approximate the correlation function by a $\delta$-function 
\begin{eqnarray*}
\langle \hat{A}(t_1)\hat{A}(t_2)\rangle= 2\tau_{c}\delta(t_1 -t_2), 
\end{eqnarray*}
where $\tau_c$ is the correlation time. Substituting into 
Eq.~(\ref{eq:td_rhos}), we have
\mbox{$\rho_s(t) \simeq \rho_s(0) -\lambda^{2}\tau_{c}t[\sigma_z,[\sigma_z,
\rho_s(0)]]. $}
The state fidelity $F(t)=\langle +\vert\rho_s(t)\vert +\rangle$ is then
\begin{eqnarray}
F(t)\simeq 1 - 2\lambda^{2}\tau_{c}t, 
\label{eq:linear_decay}
\end{eqnarray}
which corresponds to a linear
decay. It is a well known fact QZE is not observable in this 
regime \cite{misra}.

In the opposite limit of a slowly fluctuating environment, where  
$| t_1-t_2| \ll \tau_c$,
the correlation function is constant,
\begin{eqnarray*}
\langle \hat{A}(t_1)\hat{A}(t_2)\rangle  = 1, 
\end{eqnarray*}
and Eq.~(\ref{eq:td_rhos}) becomes
\mbox{$\rho_s(t) \simeq \rho_s(0) -\frac{1}{2}\lambda^{2}t^{2}
[\sigma_z,[\sigma_z, \rho_s(0)]]$}
for $\lambda^{2}\tau_{c}t$ $\ll$ $1$, 
i.e., the decoherence quadratically starts:
\begin{eqnarray}
F(t) \simeq 1 - \lambda^{2}t^2.
\label{eq:quadratic_decay}
\end{eqnarray}
This is the regime the QZE can be observed: if $N$ sequential
projective measurements of 
$\vert +\rangle\langle +\vert$ are carried out on the system and the delay 
between the measurements is short compared to the correlation time, 
$T/N \ll \tau_c$, 
the system remains in the initial state with a probability
$P(N) = (F(T/N))^N) \simeq \left(1 -\frac{\lambda^{2}T^2}{N^2}\right)^N 
\simeq e^{-\frac{\lambda^{2}T^2}{N}}$ .
Since the exponent can become arbitrarily small for large $N$, 
the decoherence vanishes asymptotically. 

\subsection{Theoretical Background of Experiment}
\label{TBE}
We intend to show a proof-of-principle demonstration with NMR that 
QZE is employed for protecting a phase decoherence. 
We need to answer two questions: 1) how to obtain a system that shows 
a quadratic decay. 2) how to measure a quantum system without 
employing ensemble averages, unlike usual NMR experiments \cite{nmr,antizenonmr}. 

\subsubsection{Quadratic Decay.}
\label{sss_quadratic_decay}
A molecule in a solvent strongly and rapidly interacts with molecules 
of the solvent. 
These interactions are so strong and rapid that they effectively cancel 
with each other. Therefore, the effective
interaction between the molecule and the solvent is often small \cite{Abragam}.  
This phenomenon is called motional narrowing. Since $T_1$ and $T_2$ of a spin 
in the molecule become longer, peaks in its NMR spectrum become sharper. 
The total number of quantum gates that can be performed
within the coherence time is estimated 
as $T_2/ \tau_G$, where $\tau_G$ denotes the time of the gate operation.
These long $T_2$ guarantees many quantum operations 
and liquid state NMR is often employed 
to demonstrate quantum algorithms.

We introduce molecules with magnetic dipole moments, originated 
from electrons, into the solvent. These molecules 
generate strong but short range magnetic fields. They are short range 
in the sense that they decay with $r^{-3}$ or faster as a function of $r$, 
the distance. These molecules are called magnetic impurities. They also
move rapidly 
in the solvent. When a magnetic impurity passes by a molecule, 
a strong interaction between a spin in the molecule and the 
magnetic impurity changes the spin state abruptly. 
Such an interaction leads to a final
spin state uncorrelated to the initial one. Motional narrowing does 
not occur, since there are not enough magnetic impurities for their effects 
to be canceled on average. 

The interaction of the spin with the magnetic impurity is a
$\delta$ function-like in time, and thus we do not expect a quadratic decay
of the spin state. The higher density of the magnetic impurity leads to
shorter $T_1$ and 
$T_2$, but the characteristic of the state decay is always linear, like in
Eq.~(\ref{eq:linear_decay}). 

Now, let us consider a two-spin 
molecule. One of the two spins (called E) is under the influence of 
a magnetic impurity and interacts with the other spin (called S) 
through a scalar coupling 
(for simplicity, the weak coupling limit \cite{levitt} is assumed)
\begin{eqnarray}
 \mathcal{H}_{\rm J} =  J \sigma_z^{\rm S} \sigma_z^{\rm E}/4,
\label{eq:Hj}
\end{eqnarray}
where $\sigma_i^{\rm E}$ $(\sigma_i^{\rm S})$ is a Pauli operator 
acting on the spin E (S). $J$ is the strength of the scalar coupling, 
in frequency units.
The spin S interacts only with the spin E and free from the 
magnetic impurity. 
The delta function-like interaction on the spin E due to the
magnetic impurity is transformed into a slower interaction on the spin S
by the scalar coupling. In this case, $J$ and $T_1$ of the spin E determine
the correlation time. However, for the case considered in this paper,
the $J$ term is more essential in the practical range of $T_1$ of 
the spin E and the correlation time becomes of the order of  $2 \pi /J$. 

We can obtain a spin showing a quadratic decay with a two-spin 
molecule and magnetic impurities. The correlation time 
is of the order of $2 \pi /J$, usually of few milliseconds for a carbon 
and hydrogen nuclei 
combination and thus this interaction is well controllable with NMR
techniques. A similar approach has been reported, but using a series of
pulses instead of magnetic impurities \cite{kondo_a_rel}.

\subsubsection{Projective Measurements.}
\label{sss_projective_meas}
A field gradient was often employed as a projective measurement in NMR 
experiments \cite{nmr,antizenonmr}. However, it is based on a spacial 
averaging, or the ensemble nature of NMR, and thus it is considered as 
a mere simulation of a projective measurement. Therefore, we cannot 
employ a field gradient as a projective measurement in our experiment.

We have to implement a 
measurement according to its definition. A measurement of a system means  
the system is entangled to a measurement apparatus. Then, 
the measurement apparatus falls into one of its eigenstates. 
The corresponding eigenvalue is treated as the measured value \cite{Zurek}. 
According to this idea, we are able to implement 
a non-selective measurement $M_{\rm DS}$ with another qubit, 
called a device (D), as follows\,
\begin{eqnarray}
\label{eq:meas_op}
 M_{\rm DS}(\rho) &=&  
  \frac{1}{2} Ad({\mathcal E}_1, \rho) + \frac{1}{2} Ad({\mathcal E}_2, \rho),
\label{eq:ideal_m}
\end{eqnarray}
where $Ad(*,\rho) = * \rho *^{\dagger}$, 
${\mathcal E}_1 = CNOT_{\rm DS}$ and 
${\mathcal E}_2 = \sigma_y^{\rm D}  \,CNOT_{\rm DS}$. 
$CNOT_{\rm DS}$ is a CNOT gate 
whose control 
and target qubits are the device D and the system S, respectively. 
We employ $CNOT_{\rm DS}$  as an entangler, 
while $\sigma_y^{\rm D}$ causes the phase decoherence of 
the device D \cite{Nielsen}. 
This measurement operation is virtual and instantaneous. 
This procedure is effective when the system and device are not correlated 
before the measurements are performed.
Such condition is satisfied, since the device is perfectly decohered by the
operator $\sigma_{y}^{D}$ at the end of each measurement.

It is worth mentioning that we use a stochastic master
equation to describe the measurement, and such a master
equation usually requires an average of the measurement outcomes
\cite{c_time_delta}.
However, the average technique is not against the fact that our scheme does not 
rely on the nature of ensemble average. Even for a single spin, one
needs to repeat the
experiment many times, and needs to take an average of the measurement
results. For example, in \cite{nvdd}, they consider a single NV center,
and a model similar 
to ours is applied to include the fluctuating noise, using an average. Similarly, 
our calculation described here can be applied to a single spin.

\subsubsection{QZE Simulation.}
\label{sss_model_QZE_e}
We combine the ideas in \S~\ref{sss_quadratic_decay} 
and \S~\ref{sss_projective_meas} 
to design an ideal QZE simulation. We consider a three-qubit system. 
The first qubit is a system S of which phase is protected. The second
qubit E mediates and filters the random noise caused by a magnetic 
impurity to the system. The third qubit is a device D.  
 
\begin{figure}[hbt]
\centering
\caption{
\label{fig:theory_QZE}
Simulated FID signal with $p_e = 0.05$ 
by Eq.~(\ref{eq:theory_FID}) (large red) 
that shows oscillatory behaviour 
but decays in time. The data without decay (green) is the case of $p_e = 0$ 
and is shown for comparison.
The three curves without oscillatory behaviours are simulations with 
the measurement $M_{\rm DS}$.
The red curve with $\displaystyle \tau_{xy} = 1/160$ decays
slower than the black one, with $\displaystyle \tau_{xy} = 1/40$, as expected. 
The blue curve (dots) with $\displaystyle \tau_{xy} = 1/10$ decays 
even faster than the envelope of the simulated FID signal. 
 }
\vspace{2ex}
\includegraphics[width=10cm]{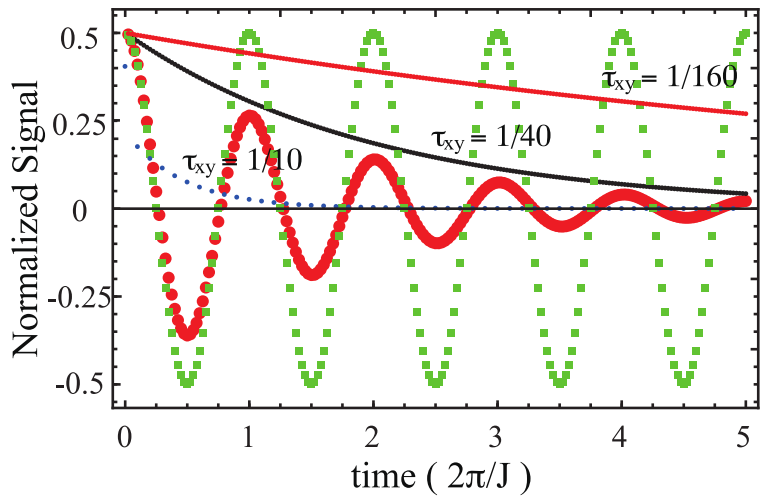}
\end{figure}

\begin{figure}[hbt]
\centering
\caption{
\label{fig:quadratic}
The differences 
between simulations 
with $p_e = 0.05$ (red) and $0.00$ (green) in the initial stage of 
the dynamics, shown in figure~\ref{fig:theory_QZE}. 
A quadratic behaviour is  observed. 
 }
\vspace{2ex}
\includegraphics[width=10cm]{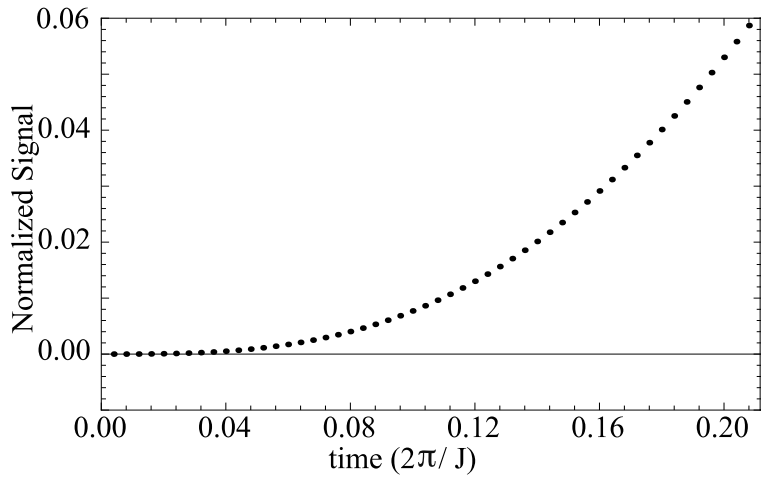}
\end{figure}

First, we consider a time development without $M_{\rm DS}$ 
as follows. 
\begin{eqnarray}
\label{eq:theory_FID}
\left[\tau_{xy} \right]^N, 
\end{eqnarray}
where $\tau_{xy}$ is the period when the system decoheres 
and $N$ is the number of repetitions. We describe the open system dynamics with 
the operator sum formalism  \cite{klaus}. 
The density matrix change during $\tau_{xy}$ is 
\begin{eqnarray}
\label{eq:theory_FID_2}
\hspace{-6ex}
\rho  \rightarrow 
  (1-p_e \tau_{xy}) Ad(e^{-i {\mathcal H}_{\rm J} \tau_{xy}}, \rho) 
+    p_e \tau_{xy}  
Ad\left( (\sigma_x^{\rm E} \cos \theta +\sigma_y^{\rm E}\sin \theta) 
e^{-i {\mathcal H}_{\rm J}  \tau_{xy}}, \rho \right),  
\end{eqnarray}
where $p_e$ determines the strength of 
decoherence, $\sigma_i^{\rm E}$ is a $\sigma_i$ operator acting on the
spin E, and $\theta$ is a random variable indicating the rotation axis of 
the $\pi$-rotation on the spin E, caused by the magnetic impurity.  
$p_e$ is a function of the concentrations of the magnetic impurity and the
molecule of interest. 
Since the initial state 
of E is $\vert 0 \rangle \langle 0 \vert_{E}$  and $H_J$ contains
only $\sigma_z^{\rm E}$, 
$Ad((\sigma_x^{\rm E} \cos \theta +\sigma_y^{\rm E}\sin \theta)^n,
\vert 0\rangle \langle 0 \vert_{E})$ is equivalent to 
$Ad((\sigma_y^{\rm E})^n,  \vert 0\rangle \langle 0 \vert_{E}) $
for the interaction $H_J$, where $n$ is an arbitrary integer.
Therefore, we can replace the second term of 
$\sigma_x^{\rm E} \cos \theta +\sigma_y^{\rm E}\sin \theta$
by $\sigma_y^{\rm E}$. 
We show the simulation results (red and green) with $p_e = 0.05$ and $0.00$
in figure~\ref{fig:theory_QZE}. The initial state is 
$\vert + \rangle \langle + \vert_{S} \otimes \vert 0 \rangle \langle 0 \vert_{E} 
\otimes \vert 0 \rangle \langle 0 \vert_{D}$. 
figure~\ref{fig:quadratic} shows the differences between simulations 
of $p_e = 0.05$ (red) and $0.00$ (green) in the initial stage of 
the dynamics, shown in figure~\ref{fig:theory_QZE}. 
A quadratic behaviour is observed. 

Second, we introduce $M_{\rm DS}$ as follows. 
\begin{eqnarray}
\label{eq:theory_QZE}
\left[\tau_{xy} - M_{\rm DS} \right]^N
\end{eqnarray}
We show the simulation results with $p_e = 0.05$ and 
three different $\tau_{xy}$'s in figure~\ref{fig:theory_QZE}.  
The result with $\displaystyle \tau_{xy} = 1/40$ (black) 
decays faster than that of $\displaystyle \tau_{xy} = 1/160$ (red),
as expected.  The result with $\displaystyle \tau_{xy} = 1/10$ decays 
even faster than the envelope of the simulated FID. 

We can obtain the same results of figure~\ref{fig:theory_QZE} with $M_{ES}$, 
where $E$ is the spin mediating the random noise.
$M_{\rm ES}$ is obtained replacing  
$CNOT_{\rm DS}$ and  
$\sigma_y^{\rm D}$ (equivalent with 
$\sigma_x^{\rm D} \cos \theta +\sigma_y^{\rm D}\sin \theta$)
in Eq.~(\ref{eq:ideal_m}) with 
$CNOT_{\rm ES}$ and $\sigma_x^{\rm E} \cos \theta +\sigma_y^{\rm E}\sin
\theta$ (equivalent with $\sigma_y^{\rm E}$), respectively. 
It implies that we are able to measure the system S with the spin E, 
without introducing the device (third spin) D.


\section{Experiments and Simulations}

\subsection{Sample}
\begin{figure}[hbt]
\centering
\caption{
\label{fig:fid_sim}
Measured (large red) and simulated (small black)
FID signals of $^{13}$C.  The $^{13}$C atom is in the center of the schematic 
chloroform molecule, surrounded by the $^{1}$H and the three chlorine atoms.
The selected parameters are $(T_d, 1/p_{s})=(6.5,300)$~ms. 
$T/N_t = 0.1$~ms in Eq.~(\ref{eq:free_dev}) is taken for the computation. }
\vspace{2ex}
\includegraphics[width=10cm]{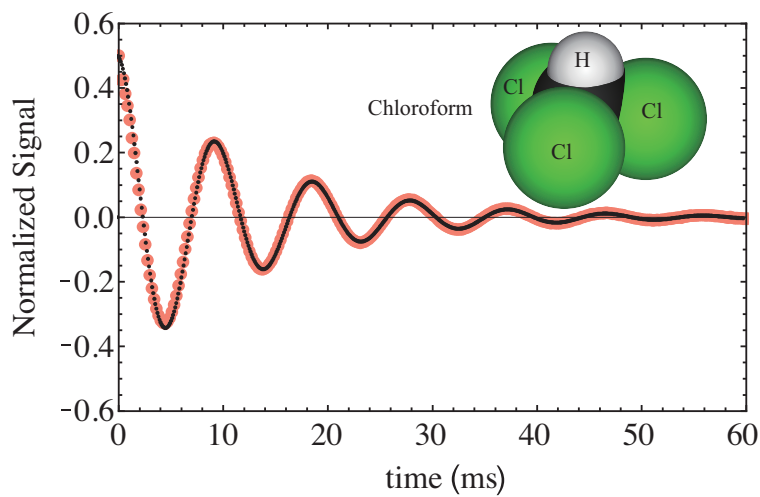}
\end{figure}

Our spectrometer is a JEOL ECA-500. 
The experiments were carried out using a $^{13}$C-labeled 
chloroform (Cambridge Isotopes) diluted in $d_6$-acetone
with a 303~mM concentration, at room temperature. 
We added a magnetic impurity (47.7~mM of Iron(III) acetylacetonate, 
a relaxation agent to control $T_1$ and $T_2$ in NMR experiments)
to the solution, which mainly introduces a random flip-flop 
motion of the $^{1}$H spin, since the carbon atom     
is surrounded and protected by three chlorine and one $^{1}$H atoms, 
as illustrated in the inset of figure~\ref{fig:fid_sim}.  
This is a realization of the configuration discussed 
in \S~\ref{sss_quadratic_decay}.

\subsubsection{$T_1$ and $T_2$ Measurements and Bang-Bang Control}
\label{sample}
To characterize the sample, we measured $T_1$ and $T_2$ of 
the $^{1}$H and $^{13}$C spins. $T_1$ and $T_2$ of the $^{1}$H are 
both about 7~ms. It is reasonable because the magnetic impurity 
simultaneously flips spins and destroys their phase coherence.
On the other hand, $T_1$ of the $^{13}$C spin is 300~ms, while its
$T_2$ is 17~ms. Without the magnetic impurities, $T_1$ and $T_2$ 
of the $^{13}$C spin are 20~s and 0.3~s 
\cite{kondo_acc}. So, the $T_2$ relaxation is speed up by about 20 times. 
We measured $T_1$  using the standard inversion recovery sequence.
Due to a good field homogeneity,
$T_2$ can be obtained directly from the Free Induction Decay (FID) 
signal \cite{levitt}, see figure~\ref{fig:fid_sim}. 
This shorter $T_2$, compared with 
$T_1$, can be understood according to the discussion in
\S~\ref{sss_quadratic_decay}. For our sample, 
$J/2\pi \approx 215$~Hz.

\begin{figure}[hbt]
\centering
\caption{
\label{fig:Cflip}
Effective $T_2$'s  
of the $^{13}$C spin 
as a function of the intervals between $\pi$-pulses of the XY-4 sequence 
applied to the $^{13}$C spin.  
}
\vspace{2ex}
\includegraphics[width=10cm]{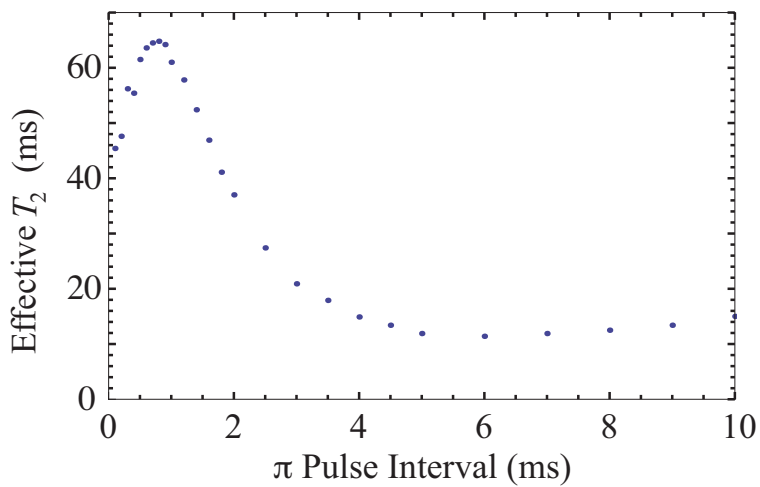}
\end{figure}

We expect a quadratic decay in the FID signal as shown in  
figure~\ref{fig:quadratic}. It is, however, 
difficult to observe such behaviour due to experimental limitations.
Therefore, we performed Bang-Bang controls on our sample, as shown 
in figure~\ref{fig:Cflip}.
Bang-Bang controls also require a quadratic 
decay to be effective and thus we are able to judge if our sample have 
it at the initial stage of decoherence.  
The effective $T_2$ of the $^{13}$C is measured through the 
application of a XY-4 sequence \cite{Mustafa,xy4}, which compensates 
pulse imperfections, to the $^{13}$C spin.  
When the interval between the pulses is below 4~ms, 
the effective $T_2$ becomes longer.
We observe a maximum in
figure~\ref{fig:Cflip} which may be understood by the discussions 
in \cite{bb_filter,opt_meas_f}. 
We will consider this elsewhere.

\begin{figure}[hbt]
\centering
\caption{
\label{fig:Hflip}
Effective $T_2$'s  
of the $^{13}$C spin 
as a function of the intervals between $\pi$-pulses of the XY-4 sequence 
applied to the $^{1}$H spin.  
}
\vspace{2ex}
\includegraphics[width=10cm]{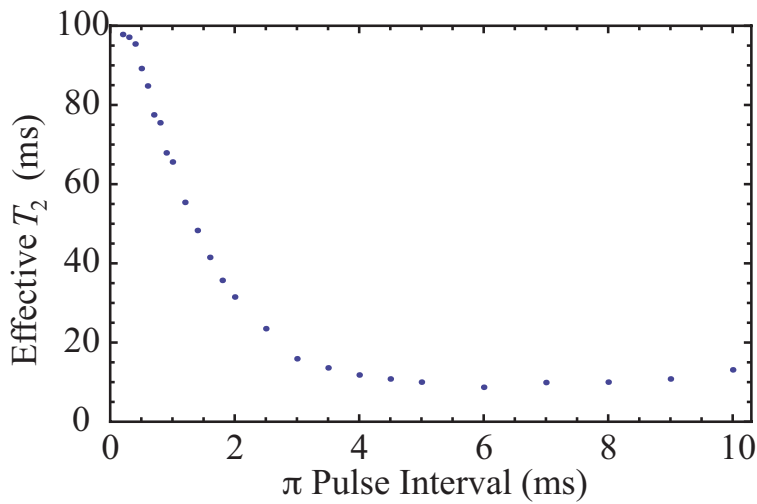}
\end{figure}

Further, we also measured this effective $T_2$ 
applying a series of $\pi$-pulses to the $^{1}$H spin, as shown in 
figure~\ref{fig:Hflip}. 
If they are frequent enough, these $\pi$-pulses 
effectively decouple the $^{1}$H and the $^{13}$C spins, 
as well known in NMR \cite{levitt}.
The effective $T_2$ at the $\pi$-pulse interval of
0.2~ms becomes $97/17 \sim 6$ times
longer than that without $\pi$-pulses. 
This result indicates that the dominant (more than 80 \%) 
source of the decoherence of the $^{13}$C spin is the $^{1}$H one.

\subsubsection{Simulation of the FID with operator sum formalism.}
\label{sss_sim_FID}
We reproduce the observed FID signal shown in figure~\ref{fig:fid_sim}
using the operator sum formalism \cite{klaus}. 

The thermal state density matrix of the two
qubit system is well approximated as
\begin{eqnarray}
\label{eq:density_th}
\rho = \left(\frac{\sigma_0}{2}\right)^{\otimes 2} 
+ \epsilon_{s}\frac{\sigma_{z}\otimes\sigma_0}{4} + 
\epsilon_{e}\frac{\sigma_{0}\otimes\sigma_z}{4}, 
\end{eqnarray}
where $\epsilon_i=\hbar\omega_{i}/2k_{B}T$, $\omega_i$ is the Larmor
frequency of the \mbox{$i$-th} spin, and $\sigma_0$ is the identity
matrix of dimension 2. 
The suffixes $s$ and $e$ denote the $^{13}$C and $^{1}$H spin, respectively. 
Eq.~(\ref{eq:density_th}) can be rewritten as 
$$\rho = (1- \epsilon_s ) \left(\frac{\sigma_0}{2}\right)^{\otimes 2} 
+ \epsilon_{s} 
 | 0\rangle \langle 0|  
\otimes \frac{\sigma_0}{2} + 
\epsilon_{e}\frac{\sigma_{0}\otimes\sigma_z}{4}. $$
Since the trace of the NMR observable, such as $\sigma_{x, y}$, is zero, 
the term $(1- \epsilon_s ) \left(\sigma_0/2\right)^{\otimes 2}$
cannot be observed in NMR experiments and can be ignored.
Moreover, only the $^{13}$C spin is observed (or,   
$Tr\left( \sigma_{x, y} \otimes \sigma_0, \rho\right)$ is measured), 
the above density matrix can be regarded as $\epsilon_s \rho_{\rm th}$, where 
$\rho_{\rm th} = | 0\rangle \langle 0|  \otimes (\sigma_0/2)$. 
This state can be normalized as a pseudopure state
$\vert 0\rangle\langle 0\vert$ for the $^{13}$C spin, 
without any initialization operation.

The effect of the $^{1}$H spin in the $^{13}$C spin dynamics during the FID  
can be explained with Eq.~(\ref{eq:theory_FID_2})
taking the $^{1}$H spin as the spin E. $p_e$ is given as 
$1/(2T_d)$ where $T_d$ is the flip-flopping time constant. 
When the evolution time $T$ is divided into $N_t$ equal steps, 
the state after the $i$-th step is obtained as 
\begin{eqnarray}
\label{eq:t2_effect}
\hspace{-6ex}
\rho_{i+1} &=& \left(1 -  p_{e} \frac{T}{N_t}\right) 
Ad(e^{-i{\mathcal H}_J T/N_t},\rho_i) 
+ p_{e} \frac{T}{N_t}
Ad\left(\left(\sigma_{0}\otimes\sigma_{y} \right) e^{-i {\mathcal H}_J T/N_t},
\rho_{i} \right). 
\end{eqnarray}

In our experiments, we also have to take into account a longitudinal relaxation of 
the $^{13}$C spin, 
a process towards the thermal state, which 
is $\rho_{th}$ in our case \cite{Nielsen}.
It can be described as

\begin{eqnarray}
\label{eq:t1_effect}
\rho_{i+1} = \left(1 - p_{s}\frac{T}{N_t}\right)
Ad(e^{-i{\mathcal H}_J T/N_t},\rho_i) 
+ p_{s}\frac{T}{N_t}\rho_{\rm th},
\end{eqnarray}
where $p_s$ is expected to be the inverse of $T_1$ of the $^{13}$C spin. 
 
When the dynamics of both the $^{13}$C and $^{1}$H spins are considered, 
the state after the $i$-th iteration is
\begin{eqnarray}\
\label{eq:free_dev}
\hspace{-10ex}
\rho_{i+1} &=& \left(1 - \left(p_{s} + p_{e}\right) \frac{T}{N_t}\right)
Ad(e^{-i{\mathcal H}_J T/N_t},\rho_i)  
+ p_{e} \frac{T}{N_t}Ad(\sigma_{0}\otimes\sigma_{y},\rho_{i})
+ p_{s} \frac{T}{N_t}\rho_{\rm th}, 
\end{eqnarray}
where 
$\left(\sigma_{0}\otimes\sigma_{y} \right) e^{-i {\mathcal H}_J T/N_t}$ 
is approximated as $\sigma_{0}\otimes\sigma_{y} $. This approximation 
is valid because we take $N_t$ such that $J T/N_t \ll 1$.

Black small dots in figure~\ref{fig:fid_sim}
show the simulated FID signal with $T/N_t = 0.1$~ms. 
Taking a proper parameter set of
$(T_d, 1/p_{s})=(6.5, 300)$~ms, our simulation reproduces 
the measured FID signal very well.
$T_d$ and $1/p_{s}$ are equal to $T_2(=T_1)$ of the $^{1}$H spin 
and $T_1$ of the $^{13}$C spin within experimental errors, respectively.

\subsection{
QZE Experiments}
We take the $^{13}$C and $^{1}$H spins as S and E 
in \S~\ref{sss_model_QZE_e}, respectively. 
We do not require a readout of the measurements, 
and thus a non-selective measurement
is sufficient to observe the QZE.
The initial state for our experiments is 
$\vert + \rangle \langle + \vert \otimes \sigma_0/2 
= Ad\left( e^{ - i \frac{\pi}{2} \sigma_y/2}
	   \otimes \sigma_0, \rho_{\rm th} \right)$,
obtained applying a $\pi/2$ pulse to the thermal state $\rho_{\rm th}$.

\subsubsection{Implementation of Non-Selective Measurements.}
\label{sss_non_sel_m}

The essential role of the $CNOT$s in $M_{\rm ES}$ and $M_{\rm DS}$ in \S~\ref{TBE} is
to entangle the two spins. So, we are able to employ another entangler,
$e^{-i\pi\sigma_{z}\otimes\sigma_{z}/4}$ \cite{kondo_tele}, which is locally equivalent to the 
CNOT employed in $M_{ES}$. 
Our measurement procedure is as follows, 
\begin{enumerate} 
\item[(i)] applying $\displaystyle e^{ \pm i \frac{\pi}{2} \sigma_y/2}
	   \otimes \sigma_0$ 
\item[(ii)] a time delay $\tau_z$ 
\item[(iii)] applying $\displaystyle  e^{\mp i \frac{\pi}{2} \sigma_y/2}
	     \otimes \sigma_0$
\end{enumerate}
and we call this sequence $M_\pm(\tau_z)$ \cite{kondo_tele}.  
The first and third steps can be almost 
ideally implemented with composite pulses 
(SCROFULOUS \cite{scr}). 
During the second step, the $^{13}$C and $^{1}$H spins are entangled 
via ${\mathcal H}_J$ and the $^{1}$H spin decoheres 
simultaneously. 
The dynamics of the $^{13}$C spin due to $M_{\pm}(\tau_z)$ is 
non-unitary, in contrast to the unitary
control used in dynamical decoupling sequences. 
If we ignore the relaxation in the step (ii), 
$M_{\pm}(\tau_z)$ is equivalent to   
\begin{eqnarray}
\label{eq:meas_ex}
\left(e^{\mp i \frac{\pi}{2} \sigma_y/2} \otimes \sigma_0\right) 
e^{- i (J \tau_z)  \sigma_z \otimes \sigma_z/4}
\left(e^{\pm i \frac{\pi}{2} \sigma_y/2} \otimes \sigma_0 \right)
= e^{\mp i (J \tau_z) \sigma_x \otimes \sigma_z/4}. 
\end{eqnarray}
This operator entangles the spins.

We need to optimize the $\tau_z$ value in our experiments, since it 
should satisfy some conflicting conditions. To maximally entangle the 
spins, we should have $\tau_z = \frac{2\pi}{2J}\approx 2.5$~ms. To fully 
decohere the $^{1}$H spin, we should have $\tau_z$ longer than $T_2$ 
for the proton spin. However, a longer $\tau_z$ reduces the QZE due 
to $^{13}$C longitudinal relaxation. 

To perform the QZE experiment, we employ the pulse sequence
\begin{eqnarray}
\label{eq:qze_m}
\left[\tau_{xy} ~ - ~ M_{\pm}(\tau_z) \right]^N, 
\end{eqnarray}
where $\tau_{xy}$ indicates a waiting time 
when the phase decoherence occurs.   
We replace $M_{\rm DS}$ in Eq.~(\ref{eq:theory_QZE}) 
with a realistic $M_\pm (\tau_z)$, which is equally effective as 
a measurement. 

\begin{figure}[hbt]
\begin{center}
\caption{
QZE experiments with $M_-(0)$ when 
$\tau_{xy}=$ (a) 0.1,  
(b) 0.2, (c) 0.3, (d) 0.5, (e) 1.0,  and  (f) 1.5~ms.  
The $x$-coordinate of the $n$'th point 
is calculated with $n(\tau_M \alpha + \tau_{xy})$,
where $\alpha$ is a fitting parameter. 
$\tau_M = 2 \times 58~\mu$s is the period of two successive 
composite pulses, while we choose
$\alpha = 0.4$ for these experiments to reproduce 
the FID signal best. 
They are compared with 
the measured FID signal (small red points). 
}\label{fig:fid_c}
\vspace{2ex}
\includegraphics[width=12cm]{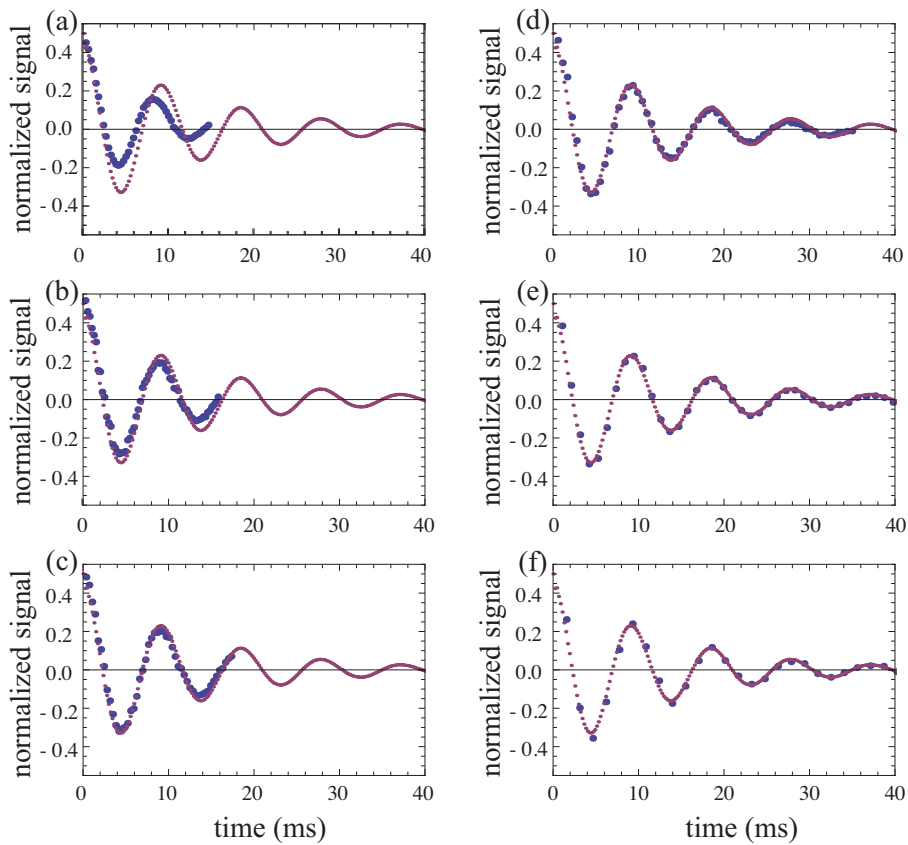}
\end{center}
\end{figure}

We confirm the importance of the period $\tau_z$ in our 
measurement ($M_\pm (\tau_z)$) taking $\tau_z = 0$ in figure~\ref{fig:fid_c}. 
It is easy to see that 
$M_\pm(0)$ is an identity operator from Eq.~(\ref{eq:meas_ex}). 
Therefore, 
we expect that an experiment of $[ \tau_{xy} - M_\pm(0) ]^N $ 
reproduces the FID signal shown in figure~\ref{fig:fid_sim}, 
if we take into account of $\tau_M$ (the period of $M_\pm(0)$ =
two successive composite pulses). 
The results with $\tau_{xy} = 0.3, 0.5, 1.0$ and 1.5~ms are identical 
with the FID signal. The results with $\tau_{xy} =$ 0.1 and 0.2 ms are 
distorted, which may be caused by the fact that our spectrometer cannot 
produce such frequent pulses.
The results described in figure~\ref{fig:fid_c}
demonstrate that the composite $\frac{\pi }{2}$ pulses do not change the decay dynamics 
induced by the proton spin, as long as no entangler is applied.

\subsubsection{QZE Experiments and Simulations.}
The simulation of the dynamics of the $^{13}$C spin 
during the QZE experiment is performed as follows. 
The pulse sequence of Eq.~(\ref{eq:qze_m}) is divided into 
$$ [\tau_{xy} - e^{ \pm i \frac{\pi}{2} \sigma_y/2}\otimes \sigma_0 
- \tau_z - e^{ \mp i\frac{\pi}{2} \sigma_y/2}\otimes \sigma_0 ]^N,$$
where $ e^{ \pm i \frac{\pi}{2} \sigma_y/2}\otimes \sigma_0 $ is 
a composite $\pi/2$ pulse acting on the $^{13}$C spin 
in the step (i) or (iii) in $M_\pm(\tau_z)$. 
The time developments during both 
$\tau_{xy}$ and $\tau_z$ are calculated using the Eq.~(\ref{eq:free_dev}). 
Additionally, the time development $(\rho \rightarrow \rho')$
during the composite $\pi/2$-pulse  with the duration $\tau = 58~\mu$s
is calculated by
\begin{eqnarray}
\label{eq:pulse_dev}
\hspace{-10ex}
\rho' &=& \left(1 - \left(p_{s} + p_{e}\right) \tau \right)
Ad(e^{-i(\mp \frac{\pi}{2} \frac{\sigma_y}{2} \otimes \sigma_0  
+ {\mathcal H}_J \tau)},\rho)  
+ p_{e} \tau Ad(\sigma_{0}\otimes\sigma_{y},\rho)
+ p_{s} \tau \rho_{\rm th}.
\end{eqnarray}
For example, in the case of $\tau_{xy} = 0.3$~ms and $\tau_z = 2.0$~ms, 
a detailed simulation procedure is as follows. The initial state of the 
experiment is $\vert +\rangle\langle +\vert\otimes\sigma_0$, which can be obtained
after the application of a $\pi/2$ pulse to $\rho_{th}$. The output 
of each step is the input of the next one.
\begin{description}
\item[a:] Simulating the time development during $\tau_{xy}$ 
with Eq.~(\ref{eq:free_dev}).
We iterate this 3 times because $\tau_{xy} = 0.3$~ms and $T/N_t = 0.1$~ms. 
\item[b:] The evolution of the state during the first rotation, 
$e^{ \pm i \frac{\pi}{2} \sigma_y/2}\otimes \sigma_0 $, is calculated using the 
Eq.~(\ref{eq:pulse_dev}).
\item[c:] The time development during $\tau_z$ is simulated with 
Eq.~(\ref{eq:free_dev}). Since $\tau_z = 2$~ms, this is iterated 20 times. 
\item[d:] The effects of the second rotation, 
$e^{ \mp i \frac{\pi}{2} \sigma_y/2}\otimes \sigma_0 $, 
are taken into account using Eq.~(\ref{eq:pulse_dev}).
\end{description}
After this last step, the whole procedure is repeated $N$ times.

\begin{figure}[hbt]
\begin{center}
\caption{
Measured (large red) and simulated QZE experiments (small black) with 
$\tau_{xy}=$ (a) 0.1,  
(b) 0.2, (c) 0.3, (d) 0.5, (e) 1.0,  and  (f) 1.5~ms 
in the case of $\tau_z = 1.0$~ms. 
The $x$-axis indicates $n$'th measurement. 
We take $M_{-}(\tau_z)$ as a  ``measurement'' in Eq.~(\ref{eq:qze_m}). 
$(T_d, 1/p_{s})=(6.5, 300)$~ms. $T/N_t = 0.1$~ms is taken for 
simulations during $\tau_{xy}$ and $\tau_z$. 
}\label{fig:qze_xx10}
\vspace{2ex}
\includegraphics[width=12cm]{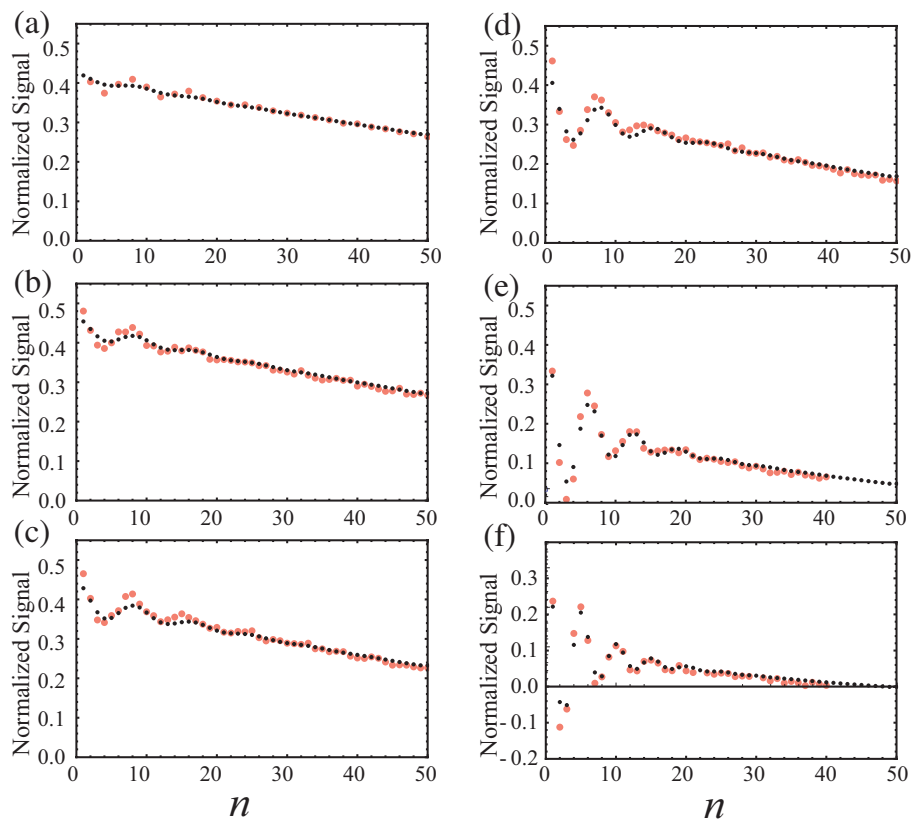}
\end{center}
\end{figure}

Figure~\ref{fig:qze_xx10} shows the results of 
the QZE experiments with $M_-(\tau_z)$ varying the parameter 
$\tau_{xy}$, but keeping a constant $\tau_z = 1.0$~ms. The data is 
compared to the simulations to check the validity of our model.  
We show the results with $M_-(\tau_z)$. 
Since the signal changes with $M_-(\tau_z)$ is larger than those with $M_+(\tau_z)$,
they should be more suitable to check the quality of the simulations. 
Our simulations reproduce the experimental results well.  

The results with $\tau_{xy} \le 0.2$~ms 
in figure~\ref{fig:fid_c} are largely distorted from the FID signal. 
On the other hand, a more oscillatory behaviour is observed for large
$\tau_{xy}$ values, as seen in figure~\ref{fig:qze_xx10}. These 
oscillations are originated from the $J$-coupling and the large 
oscillations indicate that the QZE is less effective.
Therefore, $\tau_{xy} = 0.3$~ms is the optimal measurement frequency 
for the QZE demonstration  with our imperfect measurement 
of $M_-(1.0~{\rm ms})$ \cite{opt_meas_f}.

\begin{figure}[hbt]
\begin{center}
\caption{
Measured (large red) and simulated QZE experiments (small black) with 
$\tau_z=$ (a) 0.8,  
(b) 1.0, (c) 1.2, (d) 1.5, (e) 2.0,  and  (f) 2.5~ms 
in the case of $\tau_{xy} = 0.3$~ms. 
The $x$-coordinate of the $n$'th point 
is calculated from  $n\tau_{xy}$ without taking into account 
the measurement times $\sim n\tau_{z}$.  
$(T_d, 1/p_{s})=(6.5, 300)$~ms. $T/N_t = 0.1$~ms is taken,  
while the pulse duration is  58~$\mu$s. 
The solid lines are simulations when $T_{1s} =\infty$. 
The absolute values of the measured FID signals ($\times$'s) 
are also plotted for comparison.
}\label{fig:qze_sim}
\vspace{2ex}
\includegraphics[width=\textwidth]{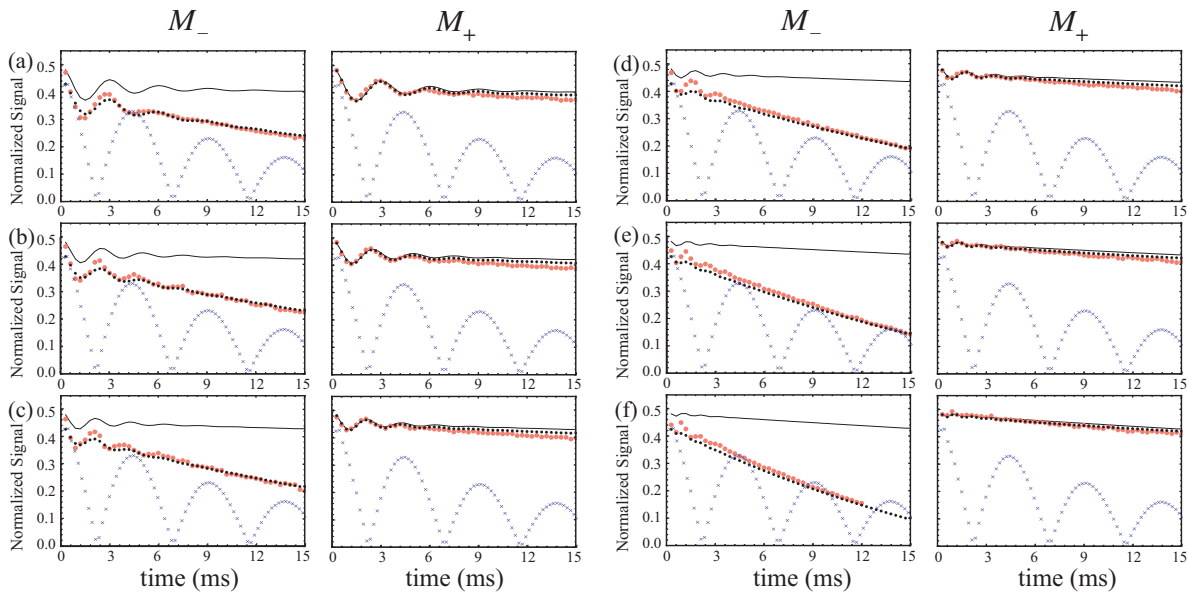}
\end{center}
\end{figure}

We vary $\tau_z $ from 0.8, 1.0, 1.2, 1.5, 2.0, and 2.5~ms in $M_\pm(\tau_z)$
with the constant $\tau_{xy} = 0.3$~ms, as shown in figure~\ref{fig:qze_sim}. 
The simulations with $M_{+}(\tau_z)$ are very similar to those 
with $T_{1s}=\infty$ (without longitudinal relaxation of the $^{13}$C spin).
It can be understood as follows. 
The state during $\tau_z$ is very close to $\rho_{\rm th}$ in the case of $M_+(\tau_z)$. 
The longitudinal relaxation given as Eq.~(\ref{eq:t1_effect}) 
is not effective in this case since the state 
during $\tau_z$ are almost the same as $\rho_{\rm th}$.
On the contrary, the state during $\tau_z$ is very close to 
$|  1 \rangle \langle 1 | \otimes \sigma_0/2$ in the case of $M_-(\tau_z)$. 
This process transforms the state 
$|  1 \rangle \langle 1 | \otimes \sigma_0/2$ to 
$|  0 \rangle \langle 0 | \otimes \sigma_0/2 = \rho_{\rm th}$, 
which means the relaxation is more effective in this case.
We are able to avoid most of the unwanted effects of longitudinal 
relaxation on the $^{13}$C spin dynamics during $\tau_z$ 
taking $M_+(\tau_z)$ as a measurement. 
The longer $\tau_z$ leads to fewer oscillations of the signal,
as seen in figure~\ref{fig:qze_sim}. This shows that $\tau_z \sim 2.5$~ms
is long enough to implement the non-selective measurement.

The signals with $M_+(\tau_z)$ decay slower than the envelope of 
the FID signal, as shown in figure~~\ref{fig:qze_sim}. 
This fact shows that the dephasing of 
$^{13}$C spin is reduced by the QZE, 
since the decay of the FID signal is caused by the 
dephasing of the $^{13}$C spin, as discussed in \S~\ref{TBE}.
It is worth mentioning that, the only difference between the figures~\ref{fig:fid_c} 
and~\ref{fig:qze_sim} are the existence of the entangling operations 
(measurements by the device), and the suppression of the decoherence in the
figure~\ref{fig:qze_sim} comes from the implementation of the measurements.
Further, a single set of parameters 
$(T_d, 1/p_s) = (6.5, 300)$~ms for the simulations can reproduce 
all the experimental results well, as shown in figures~\ref{fig:fid_sim},
\ref{fig:qze_xx10}, and \ref{fig:qze_sim}.

\section{Conclusion and Discussions}
We successfully suppressed the dephasing of an ensemble of
spins through the application of sequential 
non-selective measurements. Note that these measurements does not
rely on the ensemble nature of NMR. This is a proof-of-principle
demonstration of the Quantum Zeno Effect suppressing non-unitary 
evolution. 

One interesting extension of this work would be
the protection of an arbitrary unknown 
state, as discussed in the early stages of 
quantum information processing~\cite{nikkei}.
Here, we propose a scheme for a specific case, when 
$|  \psi_i \rangle  = \alpha |  0 \rangle + \beta |  1 \rangle$ 
is protected. 

We encode $|  \psi_i \rangle \langle \psi_i | $ to a two-qubit state 
by adding an ancillary qubit, as
\begin{eqnarray}
\rho = | \psi \rangle \langle \psi | , 
\end{eqnarray} 
where, 
$| \psi \rangle = \alpha |  ++ \rangle + \beta |  -- \rangle$.

The decoherence is described with the operator sum formalism as
\begin{eqnarray}
 \rho' &=& (1-\epsilon)\rho 
        + \frac{1}{2}\epsilon | \psi_d \rangle \langle
        \psi_d| +\frac{1}{2}\epsilon | \psi'_d \rangle \langle \psi'_d| ,
\end{eqnarray}        
where $| \psi_d \rangle$ $=$ $\alpha | -+\rangle + \beta | +-\rangle$ and
$| \psi'_d \rangle$ $=$ $\alpha | +-\rangle + \beta | -+\rangle$.

Since the error rate $\epsilon $ is supposed to be small, we ignore the terms of order
$\epsilon ^2$ or higher. We assume a quadratic decay
of the first qubit, or $ \epsilon =  \Gamma^2 t^2$, where $t$ is the
interval between the measurements. Then, we perform the following projection operator 
\begin{eqnarray}
 P_{\rm even} &=& |  ++ \rangle \langle ++ |  + |  -- \rangle \langle -- |  
\end{eqnarray}
and we obtain 
\begin{eqnarray}
 \rho'' &=& (1-\epsilon)\rho, 
\end{eqnarray}
although realization of $P_{\rm even}$ is an experimental challenge. 
One possible implementation of $P_{\rm even}$ is to use a third qubit. 
If we measure nonselectively this third qubit, after we entangle 
the third qubit with the first and second qubits, 
this  provides us with a parity projection.
When we repeat this procedure $N$-times in the period $T$, then $\rho_T$
at the time $T$ is 
\begin{eqnarray}
 \rho_T &=& \left(1-\Gamma^2 \left(\frac{T}{N} \right)^2 \right)^N \rho.
\end{eqnarray}
In the limit of infinite $N$, we obtain $\rho_T = \rho$. $\rho_T$ may
be decoded to obtain $| \psi_i \rangle$ back.

\section*{Acknowledgments}
We would like to thank Dieter Suter and Jingfu Zhang for participating 
in fruitful discussions and in the initial stages of this work. 
YK would like to thank Fumiaki Shibata, Chikako Uchiyama, and Aya Furuta 
for providing useful information. YK would like to thank partial supports of 
Grants-in-Aid for Scientific Research from JSPS (Grant No.~25400422). 
JGF thanks the Brazilian Funding agency CNPq (Grants No. PDE 236749/2012-9 and 300121/2015-6).
YM would thank a support from JSPS KAKENHI Grant
No. 15K17732.

\end{document}